\documentclass[unnumsec,webpdf,contemporary,large]{oup-authoring-template}%

\newif\ifarxiv
\arxivtrue 

\graphicspath{{Fig/}}

\theoremstyle{thmstyleone}%
\theoremstyle{thmstyletwo}%
\theoremstyle{thmstylethree}%

\usepackage{tikz}
\usetikzlibrary{calc,decorations.pathreplacing}
\usepackage{graphicx}
\usepackage{makecell}
\usepackage{multirow,array}
\usepackage{booktabs}
\usepackage{tabularx}
\usepackage{array}
\usepackage{threeparttable}
\usepackage{array}
\usepackage{pdflscape}
\usepackage{ragged2e}
\newcolumntype{L}[1]{>{\RaggedRight\arraybackslash}p{#1}}
\newcolumntype{Y}{>{\RaggedRight\arraybackslash}X}

\newcommand{\effectcell}[1]{%
  \begin{minipage}[t]{\linewidth}
  \raggedright
  #1
  \end{minipage}%
}

\newcommand{\answerline}{%
  \par\vspace{0.6\baselineskip}
  \noindent\makebox[0.55\linewidth]{\hrulefill}
  \par
}

\usepackage{amssymb}
\usepackage{array}
\usepackage{tabularx}
\usepackage{booktabs}

\newcommand{\checkbox}{$\square$}

\newtheorem*{MRQ}{Main Reasearch Question}
\newtheorem{RQ}{Reasearch Question}

\newcommand{\papertitle}{%
Comparing LLM-Based Conversational and Graphical Interfaces for Industrial Decision Tasks:
An Exploratory Mixed-Methods Study%
}

\newcommand{\papershorttitle}{LLM-Based Conversational vs Graphical Interfaces}

\newcommand{\paperabstract}{%
 The use of Generative AI Conversational User Interfaces (CUI) as a new way to access and analyze data is growing in all sectors, and the industrial one is no exception. There, large amounts of data produced by IoT devices are flowing through user interfaces and may require them a new adaptation to the new analyses needs of decision-makers. LLM-based CUIs are promising a new way to directly interact with those data through the directness of natural language and without the learning costs that every GUI design has. Moreover, the capabilities of LLMs and their agency open up the possibility to automate some tasks and help with the reasoning during decision-making activities. But are this promises well founded? We try to scope this general question with a mixed-approach study comparing a state-of-the-art dashboard with a conversational agent. A total of 20 participants used both interfaces to complete four simulated industrial decision tasks of varying complexity. We combined measures of mental workload, completion time, and decision accuracy with a post-study questionnaire and semi-structured interviews analyzed through thematic analysis. The findings suggest that the conversational agent can reduce interactional effort by supporting more direct access to information, while the dashboard remains valuable for overview and verification. However, these benefits may vary across tasks and require validation through larger-scale studies.

}

\newcommand{\paperkeywords}{%
Conversational User Interfaces, Graphical User Interfaces, Large Language Models, Conversational Agents, Dashboards, Decision Support%
}

\begin{document}

\ifarxiv


\pagestyle{plain}

\makeatletter
\twocolumn[
\begin{@twocolumnfalse}

\begin{center}

{\Large\bfseries \papertitle\par}

\vspace{1.1\baselineskip}

{\large
Roberto Figliè$^{1*}$,
Simone Caputo$^{2}$,
Alan Serrano$^{3}$,
Tommaso Turchi$^{1}$,
Daniele Mazzei$^{1}$\par
}

\vspace{0.8\baselineskip}

{\small
$^{1}$Department of Computer Science, University of Pisa, Pisa, Italy\\
$^{2}$San Matteo Hospital, Pavia, Italy\\
$^{3}$Department of Computer Science, Brunel University of London, London, United Kingdom\par
\vspace{0.5\baselineskip}
$^{*}$Corresponding author:
\href{mailto:roberto.figlie@phd.unipi.it}{roberto.figlie@phd.unipi.it}
\par
}

\end{center}

\vspace{1.2\baselineskip}

\noindent\textbf{Abstract.}
\paperabstract

\vspace{0.8\baselineskip}


\vspace{1.6\baselineskip}

\end{@twocolumnfalse}
]
\makeatother

\fi







\section{Introduction}

In industrial contexts, decision support is commonly mediated through graphical, dashboard-based interfaces \citep{lindner_behavioral_2025}.
The advent of smart manufacturing and Industry 4.0 has greatly expanded the volume and heterogeneity of data available to decision support systems, but this growth has largely been absorbed within the same visual interaction paradigm \citep{allen_data_2021}. However, in data-driven decision-making contexts, dashboards can become demanding because users must visually search, compare, and integrate information distributed across multiple elements and views \citep{van_berkel_impact_2024}. This interaction can increase mental workload by taxing limited attentional and working-memory resources, particularly when information is dense, cluttered, or not well integrated. In such conditions, information overload becomes a recurrent risk, as the amount of information available can exceed what decision makers can efficiently process and use \citep{ke_effect_2023, burnay_business_2024}.

More recently, Large Language Models (LLMs) have emerged as a promising new direction for decision support \citep{handler_large_2024}. Their appeal lies in enabling conversational access to information and analytical systems, which may reduce some of the search and integration effort otherwise required by dashboard-based interaction \citep{kernan_freire_knowledge_2024}. This trend is visible in industrial contexts as well, where recent studies have begun to explore LLM-based conversational interfaces as a way to make decision-support tools more accessible through natural-language interaction \citep{elbasheer_natural_2025,cimino_integrating_2025}. In turn, this has led to renewed interest in conversational user interfaces (CUIs) as a promising interaction modality for accessing information and system functionality \citep{deloitte_state_2026}. While CUIs are not new, recent LLMs have substantially expanded their scope and revived attention to them as a general interaction paradigm, particularly as Conversational Agents (CAs) \citep{schobel_charting_2024}. Although CUIs may still be embedded within graphical shells, they entail a different interaction logic from traditional GUIs, relying on sequential natural-language exchange and different interaction metaphors \citep{lee_exploring_2026}.

Yet whether these differences translate into better decision support remains unclear. On the one hand, conversational interaction mediated by generative AI may reduce some of the navigation and information integration effort required by visual interfaces. On the other hand, dashboards provide persistent visual overview and direct inspectability of multiple information elements. This suggests that the relative merits of CUIs and dashboards may depend less on interface novelty per se than on the cognitive and interactional demands of the task.

Despite growing interest in LLM-based conversational systems, direct empirical comparisons between CUIs and dashboards in industrial decision-support contexts remain limited. In particular, little is known about how these two interaction paradigms differ in terms of mental workload and task performance across tasks with different demands. To address this gap, we conducted an exploratory mixed-methods study comparing an LLM-based conversational interface and a dashboard in industrial decision tasks, focusing on users’ experiences and preferences while also examining mental workload and performance.

\section{Related Work}

\subsection{Dashboards as the visual interface for industrial DSS}

Dashboards have become a common visual layer through which many decision support systems expose operational and performance data. As \cite{few_information_2006} defines it, ``A dashboard is a visual display of the most important information needed to achieve one or more objectives; consolidated and arranged on a single screen so the information can be monitored at a glance". Their core function is to visually represent relevant information --usually through graphs, charts, and other visual elements-- thereby reducing reliance on more extensive reports and highlighting what matters for decision-making. Dashboard research has long framed them as a means of amplifying cognition by condensing relevant information into a visual display that supports monitoring and action \citep{yigitbasioglu_review_2012}. In this sense, they are not merely presentation tools, but a practical interface paradigm through which organizations access and act on data.

In manufacturing and operations, visualizations are described as ubiquitous, with dashboards used by managers, engineers, and shop-floor workers for monitoring, problem solving, decision making, and strategy development \citep{lindner_behavioral_2025}, sometimes even differentiating between information needs and roles across organizational levels \citep{tokola_designing_2016}. This role has become even more salient with smart manufacturing and Industry 4.0, where dashboards and related data-visualization and business-intelligence tools have been framed as enabling technologies for digital manufacturing, because they allow high-dimensional and real-time process information to be rendered in a visually interpretable form \citep{allen_data_2021,figlie_towards_2024}. In SMEs, dashboards are positioned as a practical way to turn production data into knowledge, plans, and actions, although they must be adapted to the lower maturity and resource constraints of smaller firms \citep{vilarinho_developing_2018}.  

\subsection{Limits of dashboard interaction in data-rich decision environments}

The literature also shows that dashboard effectiveness is conditional rather than automatic. Despite their popularity, it remains difficult to specify a single dashboard design that works equally well across users and tasks, because successful dashboard use depends on appropriate visualization choices, interaction design, and guidance rather than on the mere presence of a dashboard itself \citep{yigitbasioglu_review_2012,bach_dashboard_2023,hjelle_organizational_2024}. More recent work likewise shows that dashboard users often have needs that exceed simple data consumption. In workplace settings, dashboard use is intertwined with broader “data conversations” involving analysis, communication, and narrative construction, but these activities are often hindered by fragmented tool ecosystems, reliance on others, and ad hoc workarounds \citep{tory_finding_2023}. For organizational decision makers specifically, the challenge is not only analysis but information management under uncertainty, ambiguity, and competing objectives \citep{dimara_unmet_2022}.

A recurring problem is that the interactional burden of dashboards grows with data richness, interface complexity, and user diversity. Manufacturing studies report that highly flexible dashboards with many views can overwhelm novice users and therefore require explicit onboarding and guidance \citep{cibulski_reflections_2022,stoiber_visualization_2019}. Similar socio-technical barriers appear during real organizational transitions to commercial dashboarding systems, where employees face difficulties in training, using, and creating dashboards, especially as traditional manufacturing firms become more data-driven \citep{walchshofer_transitioning_2024}. The difficulty is compounded by the fact that visual and graph literacy vary substantially across users, such that standard graphical displays cannot be assumed to be equally interpretable by everyone \citep{galesic_graph_2011,boy_principled_2014, alhamadi_modeling_2022}. 

Another concern arising from recent dashboard and visualization research is information load. Although dashboards are intended to reduce uncertainty, complexity, and the burden of interpreting data \citep{bacic_role_2012,hjelle_organizational_2024}, recent studies suggest that this promise can fail when dashboard information load becomes too high, thereby increasing demands on users’ cognitive resources and contributing to overload \citep{alhamadi_modeling_2022,ke_effect_2023}. Interface adaptations such as customization may help, but only when they are meaningfully engaged with and carefully designed \citep{alsayahani_effects_2025}.
Beyond immediate task performance, such factors can also affect adoption, since informational, representational, and non-informational loads have been shown to shape dashboard acceptance and rejection \citep{burnay_business_2024}. Overall, the literature suggests that dashboard interaction is powerful but demanding, and that its success depends on a difficult balance between richness, clarity, flexibility, and learnability.

\subsection{Conversational access to analytics: from pre-LLM NLIs to LLM-based CUIs}

Attempts to alleviate these challenges through natural-language interaction predate contemporary LLMs. The literature on visualization-oriented natural language interfaces (V-NLIs) describes natural language as a complementary input modality to direct manipulation, allowing users to interact with visualizations without relying exclusively on tool-specific operations and thereby helping them focus more directly on the analysis task \citep{srinivasan_orko_2018,shen_towards_2023}. In parallel, \citet{hoque_applying_2018} argued that users should be able to focus on their analytical questions rather than on the mechanics of operating visualization tools, leading them to propose a conversational back-and-forth NLI for interacting with visualizations. Similar motivations also appeared in industrial settings: \citet{jwo_interactive_2021} proposed a virtual assistant as a mediator between users and a smart-manufacturing dashboard, motivated by the observation that obtaining more detailed information from conventional dashboards often required additional interaction effort or human assistance.

Earlier conversational access to data and visualization nevertheless remained technologically constrained. A recent scoping review by \citet{kavaz_chatbot-based_2023} shows that chatbot-based V-NLIs have so far covered only a limited range of interaction methods, with relatively weak support for more complex interactions. At the same time, the authors argue that emerging LLMs have the potential to expand the scope of conversational visual analytics by enabling richer explanations and opening new possibilities for more complex interactions. This suggests that LLM-based CUIs may overcome some of the technological constraints that limited earlier conversational interfaces in data-intensive settings. For this reason, current industrial studies have begun to explore LLM-based conversational interfaces as a way to make decision-support tools more accessible through natural-language interaction \citep{handler_large_2024,elbasheer_natural_2025,cimino_integrating_2025}. More ambitiously, recent work increasingly frames these systems not only as conversational interfaces, but as forms of Intelligent Cognitive Assistants (ICAs) intended to support the decision process itself. In industrial and AIoT settings, ICAs have been proposed as human-centred systems that help operators cope with complexity, become more context-aware, and reduce technical workload in order to focus on the task at hand \citep{angulo_towards_2023}. Early industrial LLM-based CUI work similarly positions such systems as simplifying decision-making for managers and operators rather than merely retrieving data \citep{figlie_towards_2024,colabianchi_assessment_2024}. However, existing empirical evidence remains heterogeneous, leaving open questions about the usefulness and usability of these systems \citep{kavaz_chatbot-based_2023,marconi_assessing_2026}.

\subsection{Interface logic, task demands, and the need for comparative evidence}

The key issue, then, is not whether dashboards or conversational interfaces are better in the abstract, but how their different interaction logics align with different decision tasks. Visual interfaces offer persistent overview, simultaneous access to multiple information elements, and direct inspectability; conversational interfaces instead foreground sequential query--response exchange and allow users to express informational intent more directly. This distinction is consistent with broader work on conversational interfaces for information search, which notes that moving from graphical to conversational interaction may transform users' information-seeking behaviors, while the design space remains only partially understood \citep{liao_conversational_2020}. It is also consistent with Cognitive Fit Theory (CFT), which holds that problem-solving performance improves when the form of information representation matches the cognitive processes required by the task \citep{vessey_cognitive_1991}. In this sense, graphical and conversational interfaces may be expected to differ not only in usability, but also in how well they fit different task demands.

Prior work suggests that such differences do matter. In one of the few earlier decision-aid comparisons explicitly grounded in CFT, visual and text-based query interfaces yielded different performance patterns depending on task complexity: text-based interaction was more accurate on low-complexity tasks, whereas the visual interface performed better on high-complexity tasks and was associated with lower subjective mental workload overall \citep{speier_influence_2003}. More recent work on interactive data visualizations similarly shows that interaction technique affects correctness, confidence, perceived difficulty, and cognitive load across lookup, comparison, and relation-seeking tasks \citep{van_berkel_impact_2024}.

The conversational interface literature points to a similar conclusion: benefits are real, but not uniform. Conversational interaction can reduce learning effort and feel more natural, yet it does not straightforwardly replace graphical decision aids \citep{liu_conversational_2024}. In fact, chatbot interfaces have also been shown to impose higher cognitive load and produce lower satisfaction than menu-based interfaces in some contexts \citep{nguyen_user_2022}. Taken together, these findings caution against both technological enthusiasm and dashboard conservatism. The relevant open question is therefore task-contingent. Conversational interfaces may be advantageous when users benefit from direct retrieval, guided querying, or stepwise information access, whereas dashboards may remain preferable when users need persistent overview, explicit comparison, and self-directed inspection of multiple data points. Despite growing interest in LLM-based CUIs, direct empirical comparisons with dashboard-based interfaces in realistic industrial decision tasks remain scarce, especially when task demands differ substantially. This gap motivates a comparative evaluation centered not only on performance, but also on mental workload and user experience across different task types.

\subsection{Research Questions and Rationale}
As discussed above, LLM-based CUIs may address some limitations of traditional dashboards and, more generally, of graphical user interfaces. This possibility is especially relevant for CAs enabled by recent generative AI models, which can act as natural-language access layers over complex information environments. The rationale of the present study is that, when such systems reduce the effort required to search for, integrate, and interpret task-relevant information, they may lower the perceived interactional burden associated with data-driven decision making. In turn, this could attenuate some of the conditions under which information load becomes experienced as information overload. This is particularly relevant in smart manufacturing, where decision makers often need to operate across large, heterogeneous, and interdependent data sources.

However, it is not yet clear whether LLM-based CUIs are destined (or ready) to fully replace traditional dashboards in high-stakes, production settings. While these technologies are widely adopted by end users, industrial contexts may lag behind due to technology readiness on both the organizational and technical side, as well as concerns that are sometimes well founded and could sometimes be more perception-driven. Conversely, given the increasing scale and complexity of data that decision makers face, one may question whether dashboards are approaching their practical limits, and whether a breakthrough such as generative AI is needed to substantially improve support for data-driven decision-making.
Following these considerations we therefore ask the following Main Research Question:

\begin{MRQ}
To what extent can an LLM-based Conversational Agent mitigate information overload, relative to a traditional dashboard, during performance-management decision tasks? 
\end{MRQ}

We further specify the Main RQ into four aspects to (i) operationalise it through a study and (ii) capture complementary facets of the user experience, combining objective outcomes with participants' accounts.

A central motivation for introducing a CA is the possibility of reducing users' mental workload by offloading parts of information seeking and integration to the system. In contrast, dashboards can require users to visually search, switch views, and keep intermediate values in mind while comparing or aggregating information. At the same time, conversational interaction may introduce its own sources of workload (e.g., formulating queries, keeping track of dialogue context, and handling misunderstandings or incorrect outputs). We therefore investigate differences in perceived workload and participants' accounts of overload-related experiences:

\begin{RQ}
How do an LLM-based CA and a traditional dashboard compare in perceived mental workload during task execution, and how do participants describe experiences of information overload (or its absence) while using each interface?
\end{RQ}

Both interfaces are intended as decision-support tools, and therefore they should enable users to complete tasks effectively and efficiently. At the same time, the two modalities differ substantially in how users interact with the system: dashboards require visual navigation and manual integration across views, whereas CUIs require formulating requests in language and interpreting generated outputs, with potential ambiguity and corrections. This can result in lowered task performance. We therefore compare the interfaces on task outcomes and interaction effort:

\begin{RQ}
How do the two interfaces compare in supporting performance-management decision tasks in terms of task success/accuracy, completion time, and interaction effort?
\end{RQ}

The extent to which a CA is beneficial may depend on task demands. Information-retrieval tasks typically require locating specific values, whereas more integrative problem-solving tasks require combining multiple cues, making comparisons, and sometimes performing computations before choosing an action. It is therefore plausible that the relative advantages and disadvantages of conversational versus graphical interaction differ across these task demands (e.g., sequential dialogue versus a visually persistent, multi-view workspace). We therefore ask:

\begin{RQ}
Do the relative advantages and disadvantages of the two interfaces vary as task demands shift from information retrieval to more integrative problem solving?
\end{RQ}

Beyond task outcomes and workload, adoption depends on how users evaluate the interface. This is particularly relevant for LLM-based systems, which are comparatively new in industrial decision support and may produce different outputs for similar user requests due to probabilistic generation and contextual sensitivity. As a result, users' willingness to base decisions on the system's outputs (especially without additional checks) cannot be assumed. These properties may influence perceived usefulness and the confidence users feel during task execution, as well as their stated intention to use one interface versus the other in future situations. We therefore ask:

\begin{RQ}
How do participants evaluate the two interfaces in terms of perceived usefulness and confidence?
\end{RQ}

Given limited prior evidence in this application domain, we address these questions through an exploratory, qualitatively driven mixed-methods approach, using objective task measures to contextualise and triangulate participants' accounts.

\section{Methodology}
In this section we present the methodology of the experiment, and particularly the experimental design, its procedure, the tasks definition and how we measured their complexity.

\subsection{Participants}
Participants were recruited from two universities, one in Italy and one in the UK. Participation was voluntary and no compensation or reward was provided. A total of $N=25$ participants took part in the study; $N=5$ belonged to the pilot and were excluded from the main analyses, yielding a final sample of $N=20$. The final sample had a mean age of 30.05 years ($Mdn = 29$, $SD = 8.55$), included 11 male and 9 female participants, and was predominantly composed of students with a computer science background ($n=18$). Fifteen participants completed the experimental session in Italian and five in English. Demographic characteristics of the final sample are reported in Table~\ref{tab:participants}.

\begin{table}[t]
\centering
\caption{Demographic characteristics of the final sample ($N=20$).}
\label{tab:participants}
\begin{tabular}{ll}
\toprule
Variable & Value \\
\midrule
Age (years) & $M = 30.05$, $Mdn = 29$, $SD = 8.55$ \\
Gender & Male: 11 (55\%); Female: 9 (45\%) \\
Session language& Italian: 15 (75\%); UK: 5 (25\%)\\
Highest qualification obtained& High school: 2 (10\%) \\
& Bachelor's: 4 (20\%) \\
& Master's: 9 (45\%) \\
& PhD: 5 (25\%) \\
Field of study & Computer science: 18 (90\%) \\
& Engineering: 1 (5\%) \\
& Other: 1 (5\%) \\
\botrule
\end{tabular}
\end{table}

The study protocol was approved by the Research Ethics Committee of the Brunel University and the Bioethical Committee of the University of Pisa (approval no. 19/2025). Participants provided informed consent, could withdraw at any time without penalty, and data were stored securely in pseudonymized form on respectively the University of Pisa or Brunel University servers, in accordance with GDPR.

\subsection{Experiment Design}
\label{sec:design}
The study employed a $2 \times 2$ within-subjects factorial design. The two independent variables were \textit{Interface Type} (GUI vs.\ CUI) and \textit{Task Complexity} (Information Retrieval vs.\ Problem Solving). Interface Type was operationalized as a performance management dashboard for the GUI condition and a chatbot for the CUI condition. Task Complexity was operationalized through two task types: Information Retrieval (IR), representing the lower-complexity condition, and Problem Solving (PS), representing the higher-complexity condition. Each participant completed four tasks in total: one IR task and one PS task with each interface.

Concerning order effects, task complexity was not counterbalanced: IR tasks were always administered before PS tasks, establishing a baseline. This choice was made to avoid a carryover effect whereby completing the more demanding PS tasks first could alter participants’ experience of the subsequent IR tasks by making them comparatively trivial or less cognitively representative.

\subsection{Tasks Definition}
To operationalize the two levels of task complexity, we designed tasks according to three criteria: 
\begin{enumerate}
    \item \textbf{Contextual validity}: Each task should closely simulate plausible real-world interactions with the system, thereby enhancing ecological validity. 
    \item \textbf{Appropriateness for non-expert users}: While retaining realism, the tasks should be understandable and solvable by non-domain experts—such as the student participants involved in this study. 
    \item  \textbf{Clear evaluability}: Each task was intended to admit a single, optimal solution. By making tasks as close-ended as possible, we enabled transparent and objective evaluation of task success, avoiding reliance on subjective judgment metrics. 

\end{enumerate}
\subsubsection{Task type}
We distinguish between two classes of tasks in this study, namely Information Retrieval (IR) and Problem Solving (PS) tasks. Here we define an IR task as the one whose goal is to obtain or organize specific facts or data without requiring complex interpretation. Its nature is fact-oriented and, in this study, it was treated as showing binary correctness because each task was designed around a predefined target datum or factual answer, so the response could be evaluated as either matching or not matching the expected information.

A PS task, on the other hand, involves interpreting, analyzing or making a decision based on the information retrieved. It often requires reasoning across multiple factors and it aims to provide recommendations or strategies. Although PS tasks in decision-making are often open-ended, we constrained them to be objectively scorable by ensuring high discriminability among possible alternatives ---thus meeting the third criteria. For instance, when participants are asked to select the most efficient machine configuration, the presented options differ clearly in performance metrics (e.g., values such as 1.4, 0.9, and 4.5), allowing participants to apply a relatively unambiguous minimizing or maximizing rule. 

\subsubsection{Task complexity} 
\label{sec:task_complexity}
To assess task complexity \textit{a priori} and objectively, and thus reduce potential confounding between task demands and interface effects, we adopted the formal definition of task complexity proposed by Wood \cite{wood_task_1986}. According to this model, total task complexity results from the combination of three dimensions: \textit{Component Complexity} ($TC_1$), \textit{Coordinative Complexity}($TC_2$), and \textit{Dynamic Complexity}($TC_3$).  However, since our experimental tasks did not involve dynamic changes—i.e., there were no alterations in the set of required acts, information cues, or their relationships during task execution—we excluded $TC_3$ from the computation of overall complexity. The resulting task complexity values were 3 for the two IR tasks and 9 for the two PS tasks. The full wording of all experimental tasks is reported in Appendix~\ref{app:tasks}.

\subsection{Materials}
The study employed two interface conditions: a graphical dashboard and a CUI, both designed to support decision-making in the same simulated factory scenario.

The dashboard condition used the \textit{Zerynth} platform dashboard, an industrial monitoring and analytics environment providing access to factory KPIs and production data through conventional graphical interface elements (e.g., charts and tables). Zerynth provided access to the dashboard environment used in the study. This condition represented the visual analytics baseline against which the conversational interface was compared.

The CUI condition used a conversational system developed by the authors. The full technical architecture of the system is described in a previous publication \citep{figlie_towards_2024}. The system was based on GPT-4o and delivered through Telegram as the conversational front-end. For the purposes of the present study, the chatbot was configured to answer participants' queries by autonomously exploring the factory dataset through a set of internally developed tools supporting information retrieval, filtering, comparison, aggregation, and basic analytical operations over the available KPIs and production data.

In both conditions, participants interacted with the same underlying dataset and scenario, with each task explicitly framed within a predefined historical time window in order to avoid dynamic changes in the data during task execution (Section~\ref{sec:task_complexity}). The data were synthetic rather than collected from a real factory. However, they were designed to be realistic and coherent with the industrial context of the study, so as to support plausible decision-making tasks while avoiding the use of sensitive or proprietary operational data.

All sessions were conducted on a laptop equipped with an external keyboard and mouse. Interaction data were logged for both interface conditions: for the dashboard, interaction events were captured through a JavaScript logging script injected into the browser environment; for the CUI, conversation data were collected on the chatbot database. In addition, the full session was screen-recorded using OBS Studio \footnote{https://obsproject.com/} and audio was later transcribed for use in the qualitative analysis.

\subsection{Procedure}
\begin{figure}[h]
  \centering
  \includegraphics[width=\linewidth]{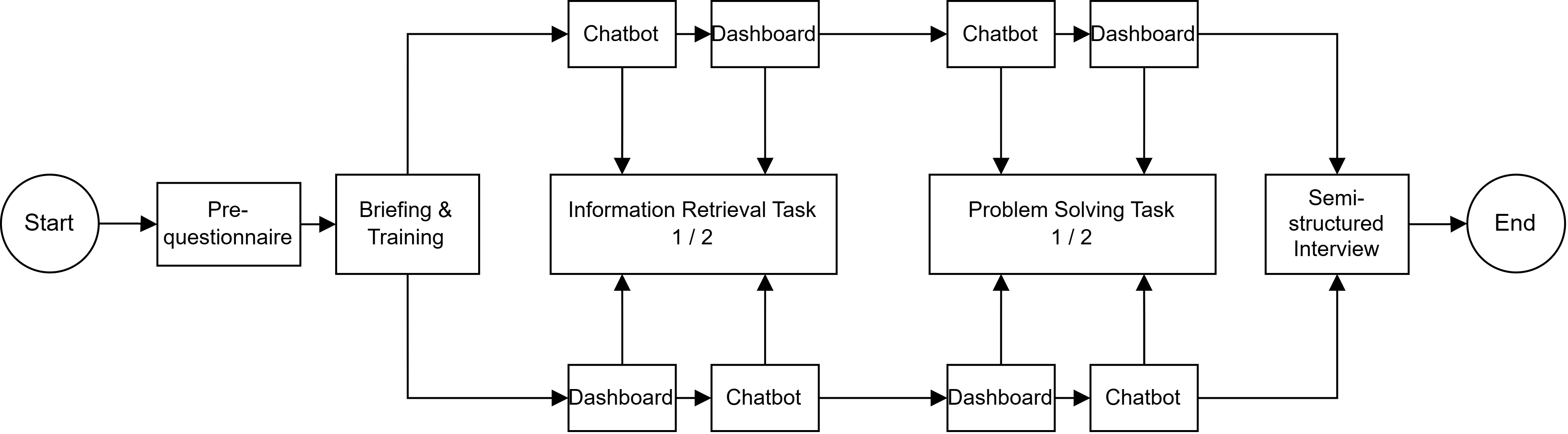}
  \caption{Diagram flow of the experimental procedure}
  \label{fig:procedure}
\end{figure}

Figure~\ref{fig:procedure} summarizes the experimental procedure. Experimental sessions were conducted between 25 March and 26 June 2025. Upon arrival, participants were welcomed, received a brief introduction to the study, and provided informed consent. They then completed a pre-questionnaire collecting background information, such as previous experience with the technologies under study and general demographics. Next, the experimenter delivered a standardized briefing about the study context, including the factory scenario and the key performance indicators relevant to the experimental tasks, without disclosing the specific tasks.

Participants were then assigned, by order of arrival, to one of four predefined order groups. These groups varied the order in which the chatbot and the dashboard were used across the task blocks. In all cases, IR tasks were completed before PS tasks, as described in Section~\ref{sec:design}. Across the session, each participant completed four tasks in total: one IR task and one PS task with the chatbot, and one IR task and one PS task with the dashboard, according to the assigned condition. Each time the participant was presented with the same dashboard starting point, while the chatbot was reset to eliminate its memory effect.

Immediately after each task, participants completed the NASA-TLX questionnaire to report their perceived mental workload during task performance. After completing all tasks, participants took part in a semi-structured interview and, in parallel, completed the related questionnaire on preferences, perceived usefulness, confidence, and interface choice in different situations. At the end of the session, participants were thanked for their participation. Each experimental session was designed to last no longer than 60 minutes.

The order of the tasks, IR before PS, was decided to avoid any learning effect between low task and high task complexity. Within each task complexity type the order is randomized.

\subsection{Qualititave Analysis Methodology}
We conducted a codebook-based thematic analysis, following the methodology outlined by \cite{guest_applied_2012}. This involved a structured process of developing a codebook and applying codes to excerpts of the interviews, to allow the identification of patterns across interviews. The codebook was refined iteratively during analysis (i.e., not treated as final until all material was coded). 

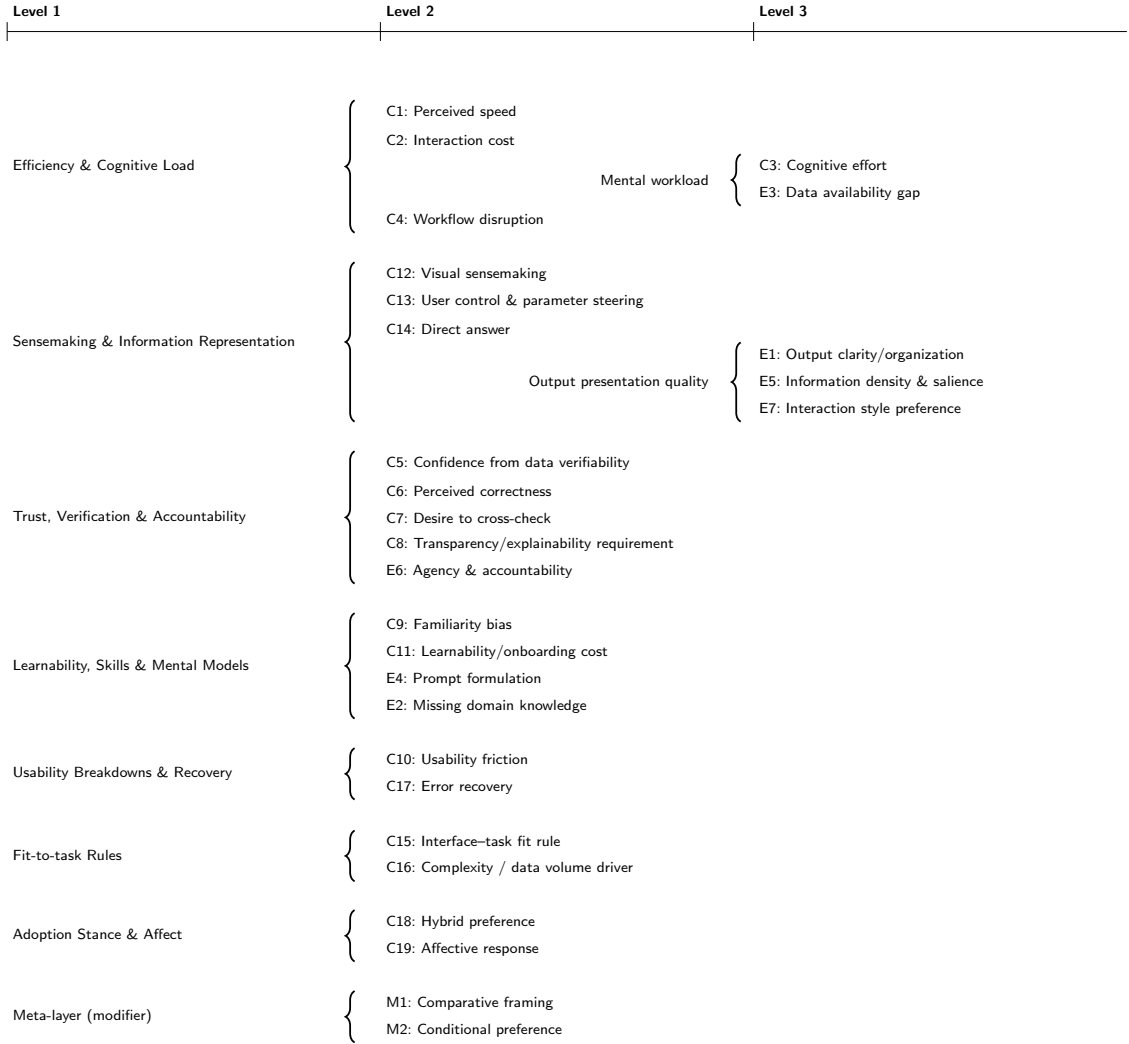
\begin{figure*}[t]
\centering
\scriptsize

\begin{minipage}{0.8\textwidth}
\centering

\resizebox{\linewidth}{!}{%
\begin{tikzpicture}[
x=1cm, y=4.2mm,
item/.style={anchor=west, align=left, text width=5.2cm},
l1label/.style={anchor=west, align=left, text width=4.4cm},
header/.style={font=\scriptsize\bfseries, anchor=south west},
brace/.style={decorate, decoration={brace, amplitude=4pt, mirror}, thick},
subbrace/.style={decorate, decoration={brace, amplitude=4pt, mirror}, thick},
subbraceLbl/.style={midway, xshift=-4mm, anchor=east, align=right, text width=3.0cm}
]

\def\xLone{0}
\def\xLtwo{5.8}
\def\xLthree{11.6}
\def\yHeader{3}

\newcommand{\LoneGroup}[4]{%
  \node[l1label] at (\xLone,#2) {#1};
  \draw[brace]
    ([xshift=-4mm]#3.north west) --
    ([xshift=-4mm]#3.north west |- #4.south);
}

\draw[thin] (\xLone,\yHeader) -- (\xLthree+5.8,\yHeader);

\foreach \x/\lab in {
  \xLone/Level 1,
  \xLtwo/Level 2,
  \xLthree/Level 3
}{
  \draw[thin] (\x,\yHeader-0.35) -- (\x,\yHeader+0.35);
  \node[header] at (\x,\yHeader+0.35) {\lab};
}

\node[item] (C1)  at (\xLtwo,   0) {C1: Perceived speed};
\node[item] (C2)  at (\xLtwo,  -1) {C2: Interaction cost};
\node[item] (C4)  at (\xLtwo,  -4) {C4: Workflow disruption};

\node[item] (C3)  at (\xLthree,-2) {C3: Cognitive effort};
\node[item] (E3)  at (\xLthree,-3) {E3: Data availability gap};

\draw[subbrace]
  ([xshift=-2mm]C3.north west) -- ([xshift=-2mm]E3.south west)
  node[subbraceLbl] {Mental workload};

\LoneGroup{Efficiency \& Cognitive Load}{-2}{C1}{C4}

\node[item] (C12) at (\xLtwo,  -6) {C12: Visual sensemaking};
\node[item] (C13) at (\xLtwo,  -7) {C13: User control \& parameter steering};
\node[item] (C14) at (\xLtwo,  -8) {C14: Direct answer};

\node[item] (E1)  at (\xLthree,-9)  {E1: Output clarity/organization};
\node[item] (E5)  at (\xLthree,-10) {E5: Information density \& salience};
\node[item] (E7)  at (\xLthree,-11) {E7: Interaction style preference};

\draw[subbrace]
  ([xshift=-2mm]E1.north west) -- ([xshift=-2mm]E7.south west)
  node[subbraceLbl] {Output presentation quality};

\LoneGroup{Sensemaking \& Information Representation}{-8.5}{C12}{E7}

\node[item] (C5)  at (\xLtwo,  -13) {C5: Confidence from data verifiability};
\node[item] (C6)  at (\xLtwo,  -14) {C6: Perceived correctness};
\node[item] (C7)  at (\xLtwo,  -15) {C7: Desire to cross-check};
\node[item] (C8)  at (\xLtwo,  -16) {C8: Transparency/explainability requirement};
\node[item] (E6)  at (\xLtwo,  -17) {E6: Agency \& accountability};

\LoneGroup{Trust, Verification \& Accountability}{-15}{C5}{E6}

\node[item] (C9)  at (\xLtwo,  -19) {C9: Familiarity bias};
\node[item] (C11) at (\xLtwo,  -20) {C11: Learnability/onboarding cost};
\node[item] (E4)  at (\xLtwo,  -21) {E4: Prompt formulation};
\node[item] (E2)  at (\xLtwo,  -22) {E2: Missing domain knowledge};

\LoneGroup{Learnability, Skills \& Mental Models}{-20.5}{C9}{E2}

\node[item] (C10) at (\xLtwo,  -24) {C10: Usability friction};
\node[item] (C17) at (\xLtwo,  -25) {C17: Error recovery};

\LoneGroup{Usability Breakdowns \& Recovery}{-24.5}{C10}{C17}

\node[item] (C15) at (\xLtwo,  -27) {C15: Interface--task fit rule};
\node[item] (C16) at (\xLtwo,  -28) {C16: Complexity / data volume driver};

\LoneGroup{Fit-to-task Rules}{-27.5}{C15}{C16}

\node[item] (C18) at (\xLtwo,  -30) {C18: Hybrid preference};
\node[item] (C19) at (\xLtwo,  -31) {C19: Affective response};

\LoneGroup{Adoption Stance \& Affect}{-30.5}{C18}{C19}

\node[item] (M1)  at (\xLtwo,  -33) {M1: Comparative framing};
\node[item] (M2)  at (\xLtwo,  -34) {M2: Conditional preference};

\LoneGroup{Meta-layer (modifier)}{-33.5}{M1}{M2}

\end{tikzpicture}%
}
\caption{Codebook}
\label{fig:codebook}
\end{minipage}
\end{figure*}

The initial codebook (C and M codes) was developed deductively from (i) the study’s conceptual focus on efficiency/workload, trust/verification, and usability; (ii) expected differences between dashboards and conversational interfaces (overview/steering vs direct answers); and (iii) the study design’s need to capture task-contingent interface choice rules and comparative framing. Codes were operationalized with definitions and include/exclude criteria. The hierarical structure of the codebook is represented in Figure \ref{fig:codebook}

Because interpretation is inherent in thematic coding, we implemented procedures to monitor coding consistency (e.g., double-coding a subset and resolving discrepancies), consistent with guidance that reliability and intercoder agreement become salient in team-based coding. Specifically, we followed the recommendations by \cite{combrinck_tutorial_2024} and \cite{nguyen-trung_chatgpt_2025} for integrating generative AI tools in qualititative data analysis. To ensure coding was not due to the specific training on one model only, we used 3 different LLMs: ChatGPT 5.2, Gemini 3 Pro, Claude 4.5 (all three with extended thinking/reasoning enabled). Each LLM was given the same codebook (in its first version). The  first time LLMs were prompted to freely apply codes to each interview and to generate emerging codes if needed. After this first pass, the human coder reviewed emerging codes and reapply the process to the interviews. At the same moment, the codebook was refined to include new emergent codes and stabilize the structure. The excerpts were then fixed by reviewing the differences in separating the text of the interviews between human and LLM coders, and decided to use the human's ones as they better identified the relevant passages. LLMs were then given again the excerpts with the updated codebook to reapply codes entirely. In the whole process, researcher coding was unknown to LLMs, as well other LLMs coding. 

The coding procedure involved an initial reliability check, after which code definitions and coding decisions were further reviewed to address ambiguous or inconsistently applied categories. After this refinement, the final coding was assessed using Krippendorff's $\alpha$ (total $\alpha = 0.738$). This value falls within the range that Krippendorff considers suitable for drawing tentative conclusions in exploratory research ($0.667 < \alpha < 0.800$) \citep{krippendorff_content_2004}. Given the exploratory purpose of the analysis and the use of genAI-assisted coding, this level of agreement was considered sufficient. As shown in Table~\ref{tab:agreement}, the lowest agreement was observed between Gemini 3 and all other coders, while the highest researcher--LLM agreement was obtained with GPT 5.2.

\begin{table}
\caption{Inter-rater reliability between coders}
\label{tab:agreement}
    \centering
    \begin{tabular}{lc}\toprule
         Coder& $\alpha$\\\midrule
         All coders& 0.738\\
         Researcher-GPT 5.2& 0.802\\
         Researcher-Gemini 3& 0.718\\
         Researcher-Claude 4.5& 0.764\\
         GPT 5.2-Gemini 3& 0.699\\
         GPT 5.2-Claude 4.5& 0.726\\
         Gemini 3-Claude 4.5&0.710\\ 
         \botrule
    \end{tabular}
\end{table}

\subsection{Objective Variables}
\subsubsection{Task Decision Accuracy}
Task decision accuracy score was measured in two different ways, following the distinction between IR and PS tasks introduced in Section~\ref{sec:task_complexity}. For IR tasks, performance was constrained to a right-or-wrong outcome; therefore, responses were coded as binary, with $success = 1$ and $failure = 0$. For PS tasks, accuracy was treated as a matter of decision optimality. These tasks required participants to choose among alternatives that could be more or less appropriate given the data, meaning that non-optimal responses were not necessarily incorrect in absolute terms, but increasingly distant from the best available option.

For each PS task, the optimal response was computed from the information cues specified in the task by applying simple maximization or minimization rules. In Task 3, the target was the cutting machine for which an automatic power-down feature would have the greatest impact, operationalized as the machine with the highest avoidable idle-energy cost. In Task 4, the target was the machine sequence most suitable for an urgent order, operationalized in terms of high performance and low average cycle time.

Accuracy score for PS tasks was then reported using Percent of Maximum Possible scoring \citep{cohen_problem_1999}. Scores were obtained by rescaling the relevant decision score, namely avoidable energy-cost savings for Task 3 and production efficiency for Task 4, against the minimum and maximum values obtainable among the available alternatives.

\subsubsection{Task completion time}
Completion time was obtained from the recording timestamps in which the task was handed to the participant to the one in which the participant stated to have finish it. To accurately capture only task completion time, we used both the recordings and the logs to identify the moment in which the interaction with the system started, hence allowing us to discard the task reading time.

\subsection{Subjective Variables}
\subsubsection{Mental workload}
Subjective Mental Workload (MWL) was measured via the NASA Task Load Index questionnaire (NASA-TLX) \citep{hart_development_1988} as a downstream indicator of possible information load or overload. Participants have been asked to fill it out after each task, for a total of four times per participant. We used the unweighted procedure for administering NASA-TLX (also known as raw NASA-TLX), and computed the total value by averaging its six subscales \citep{hart_nasa-task_2006}.

\subsubsection{Preferences}
During the last semi-structured interview, each question was accompanied by a written version to allow participants to reply both by filling out a questionnaire's item and explaining their choices and motivations. The full questionnaire and interview questions are reported in the \ref{app:interview}. The questions dealt with the preference over one interaction style over different dimensions: Perceived Usefulness, Confidence in the interface, Task-specific preference, and intention to use spanning 5 different circumstances. 

\section{Results}
\subsection{Thematic Analysis Findings}
\begin{figure}
    \centering
    \includegraphics[width=0.8\linewidth]{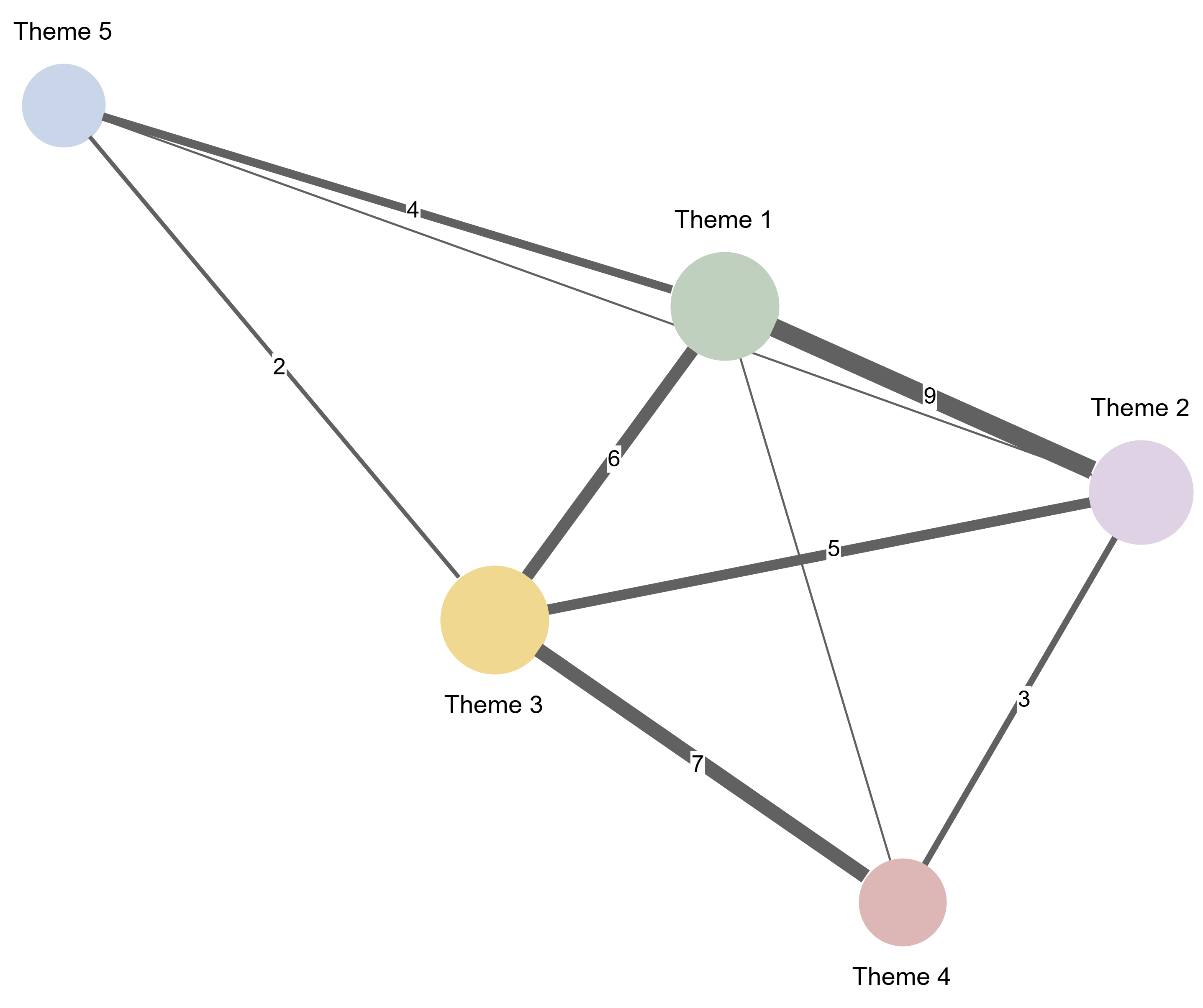}
    \caption{Theme co-occurrence network. Edge weights indicate the number of excerpts in which codes associated with two different themes co-occurred.}
    \label{fig.theme_cooccurrence}
\end{figure}

Post-session interviews were analyzed to identify how participants made sense of workload, confidence, and interface choice while completing information-retrieval and problem-solving tasks with either a CA (chatbot) or a dashboard. We report five themes that capture recurring mechanisms in these accounts and highlight the relevance of directness, reliance, skills, interaction breakdowns, and hybrid workflows. 
Moreover, quantitative outcomes (e.g., time, steps/turns, decision accuracy, NASA-TLX scores) are treated as exploratory and estimation-oriented triangulation and are integrated to check whether directions of effects are consistent with these mechanisms.
Figure~\ref{fig.theme_cooccurrence} provides a descriptive map of theme co-occurrence across coded excerpts to illustrate their interconnections. Edge weights indicate the number of excerpts in which codes associated with different themes co-occurred.

Given most of the participants conducted the study in Italian, the excerpts from their interviews are here reported translated in English for convenience.

\subsubsection{Theme 1. Conversational compression reduces interaction and mental bookkeeping}

Participants repeatedly described the CA as compressing work into a single, or generally short, exchange: ask once and receive an answer. This reduced both interaction effort and mental bookkeeping. The point was rarely speed in isolation; rather, speed was typically explained through compression and synthesis. One participant captured this perceived ease and immediacy by describing the chatbot as ``one shot, in the sense that you ask, he answers. That is, you don't even need to think about it'' (p7 e01). From this perspective, the chatbot removes intermediate steps normally required to locate, filter, and extract the needed information.

This compression was also described as a form of cognitive offloading. Participants implied that the interface does not merely retrieve a value, but also reduces the need to keep intermediate values in working memory or to perform manual comparisons and computations. In this sense, the chatbot was experienced as reducing the ``mental bookkeeping'' layer between raw data and an answer (p15 e01). When participants contrasted this with the dashboard, they often described the latter as requiring a user-managed workflow: finding the relevant view, holding values in mind, switching context, comparing values, and possibly repeating the process. A smaller set of accounts described using the chatbot not only for direct retrieval, but also for assembling or summarizing information across multiple values (p16 e01), suggesting a shift from user-produced aggregation to system-produced synthesis.

At the same time, this benefit was frequently qualified as conditional. Participants' accounts suggest that the ``one shot'' experience depended on whether the request could be formulated clearly enough for the chatbot to act on it (p13 e01; p12 e01). When this condition was not met, the interaction could become a cycle of clarification and rephrasing, reducing the advantage of conversational compression. The user-side skills underlying this condition are discussed further in Theme 3.

\subsubsection{Theme 2. Reliance is calibrated through auditability, agency, and verification}

Participants did not describe their confidence levels and reliance as a generic attitude, such as ``I trust chatbots'' or ``I do not''. Instead, they described calibrating reliance according to whether they could inspect the basis of an answer and whether responsibility for the final decision remained with them. The dashboard was often framed as supporting this form of confidence through auditability, namely the possibility of seeing the underlying values and how they relate to each other. By contrast, the chatbot was sometimes framed as creating a risk of ungrounded plausibility.

This was particularly evident when participants explicitly linked confidence to direct inspection. For example, one participant stated that ``having access to the raw data being displayed in a dashboard makes me more confident of the data being returned'' (p10 e03). In this and similar accounts, the dashboard's strength lies in supporting a defensible and verifiable chain from evidence to conclusion. Although dashboard interaction was often described as more effortful, this additional effort could also increase confidence in their responses.

Conversely, when participants experienced inconsistency or opaque reasoning from the chatbot, they described compensatory verification behaviours. Some re-asked questions to probe response stability (p21 e04), while others challenged outputs and noted changes that undermined confidence (p16 e04). The key mechanism is that the chatbot's directness and speed can be partly offset by verification labour when correctness is uncertain.

A more demanding version of this mechanism appeared when participants treated transparency as a prerequisite for reliance. For example, one participant explicitly asked about the system's method, including whether it used Retrieval Augmented Generation (RAG), as a condition for believing its answers (p14 e05). In these cases, reliance was conditional on the perceived availability of an accountable and inspectable basis for the output.

\subsubsection{Theme 3. Skill and learnability gatekeep perceived usefulness}

A recurring pattern across interviews was that participants framed both interfaces as requiring specific skills, and used these skill demands to explain the workload they experienced and their preferences. For the dashboard, several participants described an onboarding debt: being new to the interface created initial friction that was partly independent of the task itself. This effort was described as learning ``how to move'' within the interface and understanding where relevant information was located (p7 e03; p10 e02). For the chatbot, the analogous gatekeeping factor was prompting competence. Participants explicitly noted that the system performs well if users can formulate the request appropriately, and that this ability cannot be assumed for everyone (p13 e01).

Beyond interface literacy, some participants pointed to domain literacy barriers. Unfamiliarity with technical language or industrial concepts limited their ability to interpret the dashboard, especially when the interface required them to make sense of domain-specific indicators without additional explanation (p22 e02). By comparison, the chatbot was described as potentially helpful for non-domain experts because it could explain such concepts in natural language (p18 e01).

Representation comfort also appeared to matter. One participant who reported not liking graph reading explicitly preferred textual interaction (p8 e04), suggesting that the dashboard--chatbot comparison was also experienced as a contrast between visual parsing and linguistic exchange.

\subsubsection{Theme 4. Breakdowns and recovery costs undermine the promised benefits}

Participants described episodes in which interaction broke down in ways that increased effort and strongly shaped their overall evaluation of the interface. These breakdowns were often linked to recovery failures: getting lost, losing state, not knowing how to return to a previous view, or being forced into inefficient workarounds. In these accounts, perceived workload was driven less by the nominal demands of the task and more by the cost of regaining control.

A representative example concerns state loss or navigation dead ends while using the dashboard: ``I zoomed into a specific machine, I couldn't go back... it wasn't clear'' (p16 e06). Another breakdown pattern concerned comparison workflows. When participants needed to cross-check multiple machines or metrics, they described repeated reopening, sorting, and switching, which turned the task into repetitive manual work (p11 e04). A related issue arose when information was missing or not directly available. Participants described having to take notes or reconstruct values manually, increasing bookkeeping, mental calculation, and frustration (p15 e02). In some accounts, breakdowns were also framed as inadequate recovery support: the interface did not help participants get unstuck (p18 e04).

Finally, some participants described the dashboard GUI as confusing or overwhelming because of the amount of information presented. This led them to contrast it with the chatbot's clearer presentation and more direct answers. In these cases, the dashboard was experienced as inducing information load, slowing participants down and making them feel lost (p24 e01).

\subsubsection{Theme 5 (minor). Adaptive workflow orchestration under complexity and uncertainty}

A smaller but conceptually important theme was that participants sometimes rejected a binary ``best interface'' framing. Instead, they described orchestrating tools by using each interface where it was perceived to be strongest, or by imagining both interfaces as part of the same workflow.

Some participants explicitly described this complementarity: ``both at the same time... I would like to have both together'' (p10 e04), or ``one completes the other'', with the caveat that chatbot answers must be correct (p13 e03). Others articulated a more specific selection logic. Dashboards were considered useful when users did not yet know what to ask, because an overview can support discovery; the chatbot was considered useful once a specific query or synthesis goal had been formulated (p9 e03). Under higher comparison burden, the chatbot was also described as reducing the manual work of putting values ``side by side'' (p11 e03).

\subsection{Quantitative Results}
\label{sec:quant_results}
Throughout this section, tasks are labeled T1--T4 for convenience, corresponding to Task 1 through Task 4.

\subsubsection{Subjective Mental Workload}
\label{subsec:results-workload}

Workload was analyzed using rank-based tests. Independent comparisons---for example, contrasts between tasks assigned to different participants within the same interface, as well as cross-interface contrasts for the same task---were tested with Mann--Whitney U (MWU), which assesses whether values from one group tend to be larger than those from the other. Paired comparisons---that is, within-participant contrasts between IR and PS tasks within the same interface---were tested with Wilcoxon signed-rank tests, which assess whether paired differences are centered around zero. This analytic choice follows an ordinal measurement stance: Raw NASA-TLX scores were treated as bounded ordered ratings for which interval spacing cannot be assumed, and recent methodological work has recommended caution in the routine treatment of NASA-TLX as a parametric continuous measure in HCI \citep{liddell_analyzing_2018,babaei_should_2025}. 

Significant paired differences were observed for T1 vs T3 in the Dashboard condition ($p=0.0039$) and the Chatbot condition ($p=0.0020$), and for T2 vs T4 in the Chatbot condition ($p=0.0020$). Among the independent contrasts, only the cross-interface comparison in the first task was significant ($p=0.0089$); the remaining workload comparisons were not.

\begin{table*}[t]
\centering
\scriptsize
\setlength{\tabcolsep}{3pt}
\renewcommand{\arraystretch}{1.18}

\begin{threeparttable}
\centering
\caption{Workload comparisons across task and interface conditions.}
\label{tab:results-workload}

\begin{tabularx}{0.80\textwidth}{@{}L{1.8cm}L{1.8cm}L{1.8cm}L{1.8cm}L{1.8cm}Y@{}}
\toprule
Condition & Comparison & Test & Statistic & $p$ & Effect summary (95\% CI) \\
\midrule

\multicolumn{6}{@{}l}{\textit{Within-interface comparisons}} \\
\addlinespace[0.25em]

Dashboard
& T1 vs T3
& Wilcoxon
& $W=0.0$
& $\mathbf{.0039}^{*}$
& \effectcell{
$r_{rb}=-1.000$ [$-1.000$, $-1.000$]\\
HL shift $=-16.249$ [$-32.916$, $-7.500$]
} \\

Dashboard
& T2 vs T4
& Wilcoxon
& $W=13.0$
& $.1602$
& \effectcell{
$r_{rb}=-0.600$ [$-1.000$, $0.000$]\\
HL shift $=-6.250$ [$-21.666$, $5.417$]
} \\

Chatbot
& T1 vs T3
& Wilcoxon
& $W=0.0$
& $\mathbf{.0020}^{*}$
& \effectcell{
$r_{rb}=-1.000$ [$-1.000$, $-1.000$]\\
HL shift $=-22.501$ [$-27.084$, $-9.166$]
} \\

Chatbot
& T2 vs T4
& Wilcoxon
& $W=0.0$
& $\mathbf{.0020}^{*}$
& \effectcell{
$r_{rb}=-1.000$ [$-1.000$, $-1.000$]\\
HL shift $=-10.416$ [$-17.084$, $-4.167$]
} \\

\midrule

\multicolumn{6}{@{}l}{\textit{Between-interface comparisons}} \\
\addlinespace[0.25em]

T1
& Chatbot vs Dashboard
& MWU
& $U=15.5$
& $\mathbf{.0089}^{*}$
& \effectcell{
$A=0.155$ [$0.010$, $0.350$]\\
Cliff's $\delta=-0.690$ [$-0.980$, $-0.300$]\\
HL shift $=-9.167$ [$-30.834$, $-1.667$]
} \\

T2
& Chatbot vs Dashboard
& MWU
& $U=27.0$
& $.0892$
& \effectcell{
$A=0.270$ [$0.075$, $0.510$]\\
Cliff's $\delta=-0.460$ [$-0.850$, $0.020$]\\
HL shift $=-7.500$ [$-15.000$, $0.000$]
} \\

T3
& Chatbot vs Dashboard
& MWU
& $U=27.0$
& $.0892$
& \effectcell{
$A=0.270$ [$0.040$, $0.530$]\\
Cliff's $\delta=-0.460$ [$-0.920$, $0.060$]\\
HL shift $=-22.083$ [$-31.666$, $2.500$]
} \\

T4
& Chatbot vs Dashboard
& MWU
& $U=40.0$
& $.4813$
& \effectcell{
$A=0.400$ [$0.150$, $0.665$]\\
Cliff's $\delta=-0.200$ [$-0.700$, $0.330$]\\
HL shift $=-5.833$ [$-19.166$, $8.750$]
} \\

\bottomrule
\end{tabularx}
\begin{tablenotes}[flushleft]
\scriptsize
\item Note. MWU = Mann--Whitney U test; HL = Hodges--Lehmann. $r_{rb}$ denotes the rank-biserial correlation; $A$ denotes the Vargha--Delaney effect size. Asterisks indicate $p<.05$.
\end{tablenotes}

\end{threeparttable}
\end{table*}

\subsubsection{Task completion time}
\label{subsec:results-time}

Time on task was analyzed on the log scale (Table \ref{tab:results-time}). Independent comparisons were tested with Welch’s t-tests, and paired comparisons with paired-samples t-tests. Effects are reported as geometric mean ratios (GMRs), obtained by back-transforming mean log differences, together with standardized effect sizes (Hedges’$g$ for independent contrasts and Cohen’s $d_z$ for paired contrasts). For interpretation, $GMR>1$ indicates longer times in the first condition and $GMR<1$ indicates shorter times in the first condition. The largest effects were observed in the paired contrasts T1 vs T3 and T2 vs T4 in both interfaces (i.e., IR tasks vs PS tasks within interface). Additional statistically significant differences were observed for chatbot vs dashboard in task 1 and task 3 in the cross-interface comparisons.

\begin{table*}[t]
\centering
\scriptsize
\setlength{\tabcolsep}{4pt}
\renewcommand{\arraystretch}{1.15}

\begin{threeparttable}
\centering
\caption{Time on task comparisons across task and interface conditions.}
\label{tab:results-time}

\begin{tabularx}{0.73\textwidth}{@{}L{1.75cm}L{2.0cm}L{1.6cm}L{2.5cm}L{1.5cm}Y@{}}
\toprule
Condition & Comparison & Test & $p$ & GMR & Standardized effect \\
\midrule

\multicolumn{6}{@{}l}{\textit{Within-interface comparisons}} \\
\addlinespace[0.25em]

Dashboard
& T1 vs T3
& Paired $t$
& $\mathbf{.0010}^{*}$
& $0.335$
& Cohen's $d_z=-1.507$ \\

Dashboard
& T2 vs T4
& Paired $t$
& $\mathbf{.0018}^{*}$
& $0.300$
& Cohen's $d_z=-1.387$ \\

Chatbot
& T1 vs T3
& Paired $t$
& $\mathbf{.0042}^{*}$
& $0.370$
& Cohen's $d_z=-1.200$ \\

Chatbot
& T2 vs T4
& Paired $t$
& $\mathbf{7.894 \times 10^{-6}}^{*}$
& $0.250$
& Cohen's $d_z=-2.873$ \\

\midrule

\multicolumn{6}{@{}l}{\textit{Between-interface comparisons}} \\
\addlinespace[0.25em]

T1
& Chatbot vs Dashboard
& Welch's $t$
& $\mathbf{.0209}^{*}$
& $0.592$
& Hedges' $g=-1.084$ \\

T2
& Chatbot vs Dashboard
& Welch's $t$
& $.7561$
& $1.083$
& Hedges' $g=0.136$ \\

T3
& Chatbot vs Dashboard
& Welch's $t$
& $\mathbf{.0368}^{*}$
& $0.536$
& Hedges' $g=-0.966$ \\

T4
& Chatbot vs Dashboard
& Welch's $t$
& $.2058$
& $1.302$
& Hedges' $g=0.562$ \\

\bottomrule
\end{tabularx}

\begin{tablenotes}[flushleft]
\scriptsize
\item Note. GMR = geometric mean ratio. Within-interface comparisons used paired $t$ tests on log-transformed time; between-interface comparisons used Welch's $t$ tests on log-transformed time. Asterisks indicate $p<.05$.
\end{tablenotes}

\end{threeparttable}
\end{table*}

\subsubsection{Decision Accuracy Score}
\label{subsec:results-accuracy}

Decision accuracy was operationalized as a task-specific researcher-defined score. Specifically, IR tasks were scored binarily (success vs. failure), whereas PS task scoring depended on the structure of the task itself and was subsequently normalized using POMP scoring \citep{cohen_problem_1999}. Because the scoring format differed across tasks, contrasts were analyzed with tests matched to the outcome type. Independent binary comparisons were tested with two-proportion $z$ tests and Fisher's exact tests, independent continuous comparisons with Welch's $t$ tests, and paired mixed comparisons with Wilcoxon signed-rank and sign tests.

In the Dashboard condition, evidence of score differences depended on the test used. T1 vs T2 was significant under the two-proportion $z$ test ($p=0.0191$) but not under Fisher's exact test ($p=0.0573$). Likewise, T1 vs T3 was significant under the sign test ($p=0.0391$) but not under Wilcoxon signed-rank ($p=0.0781$). The remaining Dashboard comparisons were not statistically significant. In the Chatbot condition, no within-interface comparisons were statistically significant under any of the applied tests.

Across interfaces, no significant difference emerged for T1, and T2 Chatbot vs Dashboard approached but did not reach significance ($p=0.0679$ with the two-proportion $z$ test; $p=0.1698$ with Fisher's exact test). For the PS tasks, T3 accuracy scores were significantly higher in the Chatbot than in the Dashboard condition according to Welch's $t$ test ($p=0.0301$), whereas no significant difference emerged for T4 ($p=0.3072$). Overall, the only statistically clear between-interface difference in accuracy score was observed for Task 3, in favour of the Chatbot.

\begin{table*}[t]
\centering
\scriptsize
\setlength{\tabcolsep}{4pt}
\renewcommand{\arraystretch}{1.18}

\begin{threeparttable}
\caption{Accuracy score comparisons across task and interface conditions.}
\label{tab:results-accuracy}

\begin{tabularx}{0.73\textwidth}{@{}L{1.75cm}L{2.0cm}L{2.5cm}L{1.5cm}Y@{}}
\toprule
Condition & Comparison & Test(s) & $p$ & Effect summary (95\% CI) \\
\midrule

\multicolumn{5}{@{}l}{\textit{Within-interface comparisons}} \\
\addlinespace[0.25em]

Dashboard
& T1 vs T3
& \effectcell{
Wilcoxon $W=7$\\
Sign test
}
& \effectcell{
$.0781$\\
$\mathbf{.0391}^{*}$
}
& \effectcell{
$d_z=0.745$ [$0.137$, $2.229$]\\
mean diff $=0.326$ [$0.065$, $0.568$]
} \\

Dashboard
& T2 vs T4
& \effectcell{
Wilcoxon $W=3$\\
Sign test
}
& \effectcell{
$.0781$\\
$.1250$
}
& \effectcell{
$d_z=-0.737$ [$-1.609$, $-0.215$]\\
mean diff $=-0.378$ [$-0.690$, $-0.075$]
} \\

Chatbot
& T1 vs T3
& \effectcell{
Wilcoxon $W=7$\\
Sign test
}
& \effectcell{
$.2812$\\
$.1250$
}
& \effectcell{
$d_z=0.192$ [$-0.336$, $2.359$]\\
mean diff $=0.078$ [$-0.206$, $0.264$]
} \\

Chatbot
& T2 vs T4
& \effectcell{
Wilcoxon $W=13$\\
Sign test
}
& \effectcell{
$.8906$\\
$.4531$
}
& \effectcell{
$d_z=-0.207$ [$-0.735$, $0.967$]\\
mean diff $=-0.098$ [$-0.400$, $0.156$]
} \\

\midrule

\multicolumn{5}{@{}l}{\textit{Between-interface comparisons}} \\
\addlinespace[0.25em]

T1
& Chatbot vs Dashboard
& \effectcell{
Two-proportion $z$\\
Fisher exact
}
& \effectcell{
$1.0000$\\
$1.0000$
}
& \effectcell{
$RD=0.000$ [$-0.386$, $0.386$]\\
$OR=1.00$ [$0.09$, $11.42$]\\
$h=0.000$ [$-1.159$, $0.927$]
} \\

T2
& Chatbot vs Dashboard
& \effectcell{
Two-proportion $z$\\
Fisher exact
}
& \effectcell{
$.0679$\\
$.1698$
}
& \effectcell{
$RD=0.400$ [$-0.197$, $0.775$]\\
$OR=4.91$ [$0.77$, $31.32$]\\
$h=0.845$ [$0.000$, $1.982$]
} \\

T3
& Chatbot vs Dashboard
& \effectcell{
Welch $t=2.408$\\
$df=14.3$
}
& $\mathbf{.0301}^{*}$
& \effectcell{
Hedges' $g=1.031$ [$0.343$, $1.930$]\\
Cliff's $\delta=0.540$ [$0.035$, $0.900$]
} \\

T4
& Chatbot vs Dashboard
& \effectcell{
Welch $t=1.065$\\
$df=12.4$
}
& $.3072$
& \effectcell{
Hedges' $g=0.456$ [$-0.476$, $1.249$]\\
Cliff's $\delta=0.004$ [$-0.520$, $0.495$]
} \\

\bottomrule
\end{tabularx}

\begin{tablenotes}[flushleft]
\scriptsize
\item Note. RD = risk difference; OR = odds ratio. Within-interface comparisons used Wilcoxon signed-rank tests and sign tests. Between-interface comparisons used two-proportion $z$ tests and Fisher's exact tests for binary accuracy scores, and Welch's $t$ tests for continuous accuracy scores. Asterisks indicate $p<.05$.
\end{tablenotes}

\end{threeparttable}
\end{table*}

\subsubsection{Perceived usefulness, confidence, preferences, and intended interface choice}
\label{subsec:results:preferences}

Table \ref{tab:results-usefulness} reports within-subject comparisons between the chatbot and the dashboard for perceived usefulness, confidence in the interface, and task-specific preference (IR and PS, respectively). Perceived usefulness ratings were significantly higher for the chatbot than for the dashboard (one-sided Wilcoxon signed-rank test, $W=91.00$, $p=0.0038$). Descriptively, 12 participants rated the chatbot higher, 2 rated the dashboard higher, and 6 gave equal ratings. Confidence in the interface did not differ in the predicted direction: among decisive responses, 8 participants favoured the chatbot and 8 favoured the dashboard (one-sided exact binomial test, $p=0.5982$). For information retrieval tasks, decisive preferences were 8 for the chatbot and 4 for the dashboard, with 8 participants indicating both; this comparison was not significant ($p=0.1938$). For problem-solving tasks, 11 participants preferred the chatbot and 3 preferred the dashboard, with 6 indicating both; this comparison was significant ($p=0.0287$).

Table \ref{tab:results-circumstances} reports participants' intended interface choice under different circumstances. The chatbot was selected significantly more often than the dashboard when speed was the criterion (15 vs.\ 3 decisive responses; $p=0.0038$) and when a direct answer was needed (14 vs.\ 2; $p=0.0021$). No significant difference was found for task complexity (5 vs.\ 5; $p=0.6230$) or for analysis (8 vs.\ 7; $p=0.5000$). When confidence was the criterion, all decisive responses favoured the dashboard (0 vs.\ 18; one-sided $p=1.0000$ under the directional hypothesis chatbot $>$ dashboard).

\begin{table*}[t]
\centering
\scriptsize
\setlength{\tabcolsep}{4pt}
\renewcommand{\arraystretch}{1.18}

\begin{threeparttable}
\caption{Within-subject comparisons between Chatbot and Dashboard for usefulness, confidence, and task-specific preference.}
\label{tab:results-usefulness}

\begin{tabularx}{0.78\textwidth}{@{}L{3.0cm}L{3.0cm}L{1.1cm}Y@{}}
\toprule
Outcome tested & Main test statistic & $p$ & Effect summary \\
\midrule

Usefulness 
& \effectcell{
Wilcoxon signed-rank\\
one-sided, $W=91.00$
}
& $\mathbf{.0038}^{*}$
& \effectcell{
Mean(Chatbot--Dashboard) $=0.55$; median $=1.00$\\
Chatbot $>$ Dashboard: 12; Dashboard $>$ Chatbot: 2; ties: 6\\
Rank-biserial $r=0.733$ [$0.333$, $1.000$]
} \\

Interface confidence 
& \effectcell{
Exact binomial\\
one-sided, $k=8/16$
}
& $.5982$
& \effectcell{
Chatbot $=8$; Dashboard $=8$; both $=4$\\
Among decisive responses, $50.0\%$ favoured Chatbot
} \\

Preference for IR tasks 
& \effectcell{
Exact binomial\\
one-sided, $k=8/12$
}
& $.1938$
& \effectcell{
Chatbot $=8$; Dashboard $=4$; both $=8$\\
Among decisive responses, $66.7\%$ favoured Chatbot
} \\

Preference for PS tasks 
& \effectcell{
Exact binomial\\
one-sided, $k=11/14$
}
& $\mathbf{.0287}^{*}$
& \effectcell{
Chatbot $=11$; Dashboard $=3$; both $=6$\\
Among decisive responses, $78.6\%$ favoured Chatbot
} \\

\bottomrule
\end{tabularx}

\begin{tablenotes}[flushleft]
\scriptsize
\item Note. Inferential tests are one-sided tests on decisive within-subject responses only, with $H_1$: Chatbot $>$ Dashboard. Responses coded as ``both'' were excluded from inferential tests but are reported descriptively. Asterisks indicate $p<.05$.
\end{tablenotes}

\end{threeparttable}
\end{table*}

\begin{table*}[t]
\centering
\scriptsize
\setlength{\tabcolsep}{4pt}
\renewcommand{\arraystretch}{1.18}

\begin{threeparttable}
\caption{Participant-reported intended interface choice under different circumstances.}
\label{tab:results-circumstances}

\begin{tabularx}{0.77\textwidth}{@{}L{3.6cm}L{3.0cm}L{1.1cm}Y@{}}
\toprule
Circumstance tested & Main test statistic & $p$ & Effect summary \\
\midrule

When speed is important
& \effectcell{
Exact binomial\\
one-sided, $k=15/18$
}
& $\mathbf{.0038}^{*}$
& \effectcell{
Chatbot $=15$; Dashboard $=3$; both $=2$\\
Among decisive responses, $83.3\%$ favoured Chatbot
} \\

When task complexity is high
& \effectcell{
Exact binomial\\
one-sided, $k=5/10$
}
& $.6230$
& \effectcell{
Chatbot $=5$; Dashboard $=5$; both $=10$\\
Among decisive responses, $50.0\%$ favoured Chatbot
} \\

When analysis is needed
& \effectcell{
Exact binomial\\
one-sided, $k=8/15$
}
& $.5000$
& \effectcell{
Chatbot $=8$; Dashboard $=7$; both $=5$\\
Among decisive responses, $53.3\%$ favoured Chatbot
} \\

When a direct answer is needed
& \effectcell{
Exact binomial\\
one-sided, $k=14/16$
}
& $\mathbf{.0021}^{*}$
& \effectcell{
Chatbot $=14$; Dashboard $=2$; both $=4$\\
Among decisive responses, $87.5\%$ favoured Chatbot
} \\

When confidence is needed
& \effectcell{
Exact binomial\\
one-sided, $k=0/18$
}
& $1.0000$
& \effectcell{
Chatbot $=0$; Dashboard $=18$; both $=2$\\
Among decisive responses, $0.0\%$ favoured Chatbot; $100.0\%$ favoured Dashboard
} \\

\bottomrule
\end{tabularx}

\begin{tablenotes}[flushleft]
\scriptsize
\item Note. Inferential tests are one-sided exact binomial tests on decisive within-subject responses only, with $H_1$: Chatbot $>$ Dashboard. Responses coded as ``both'' were excluded from inferential tests but are reported descriptively. Asterisks indicate $p<.05$.
\end{tablenotes}

\end{threeparttable}
\end{table*}

\section{Discussion}
Rather than indicating a globally superior interface, the results point to a more conditional pattern shaped by directness, auditability, and user capabilities. In the following sections, we answer the research questions by triangulating qualitative and quantitative findings, highlighting where the two strands converge, diverge, or complement one another.

\subsection{Cognitive offloading and Information Load}

In relation to \textbf{RQ1}, the clearest point of convergence between qualitative and quantitative findings concerns what Theme 1 identified as \emph{conversational compression}. Participants repeatedly described the chatbot as most helpful when the task could be reduced to retrieval, filtering, or delegated calculation. In these cases, a sequence of operations that would otherwise require navigating the dashboard, locating the relevant view, holding intermediate values in mind, and comparing them manually could instead be compressed into a single question. This possibility of collapsing several data-handling operations into one interaction was valued by many participants and was often experienced as a form of cognitive offloading \citep{risko_cognitive_2016}, not merely as speed---although it was an appreciated related aspect. Quantitatively, this interpretation is consistent with the fact that participants reported a significantly greater intention to use the chatbot when speed was important and when a direct answer was needed.

Descriptively, subjective MWL was lower with the CA across all four tasks. This is coherent with participants' accounts that the chatbot reduced the need to search and manually assemble the answer. This pattern diverges from \citet{nguyen_user_2022}, where participants experienced higher MWL in the CUI condition than in the GUI condition. One possible explanation is that the conversational technology evaluated in that study may differ substantially from the LLM-based system used here. The system described by \citet{nguyen_user_2022} is not clearly identifiable as an LLM-based interface, and the study predates the widespread adoption of current genAI chatbots. Therefore, the divergence may partly reflect changes in conversational-system capabilities. At the same time, the workload evidence should be interpreted cautiously. Beyond the first task, cross-interface differences did not reach significance, suggesting that the perceived workload advantage of the CA, while directionally consistent with Theme 1, was not uniformly strong across all tasks in this sample. One plausible contributing factor is that part of the dashboard workload reflected not only onboarding and first-use friction, as suggested by Theme 3, but also the breakdowns captured in Theme 4. Participants described getting lost in the dashboard, struggling to combine information across views, and sometimes receiving too much information at once. These experiences could increase mental effort independently of the nominal task demand, creating a low-visual-momentum situation that has long been recognized in classic Human--Computer Interaction work \citep{woods_visual_1984,bennett_visual_2012}.

A further reason why workload differences may not have been more clearly evident is that the CA introduced a different kind of effort. When participants suspected that its answers were wrong or insufficiently grounded in the data, they described engaging in verification behaviour, for instance by asking again to probe the correctness or stability of the output (Theme 2). In this sense, the interface could feel fast while still generating additional cognitive burden when it triggered doubt. This is compatible with recent accounts suggesting that genAI can also become an unexpected source of workload \citep{ranganathan_ai_2026,yang_fatigued_2026}.

\subsection{Limits and Conditions of Conversational Support}
The qualitative sense of reduced effort associated with the CA did not translate into a uniform objective performance advantage (\textbf{RQ2}). Completion time significantly favoured the chatbot only in two of the four tasks, while in the remaining tasks the difference was not statistically clear, with the dashboard even descriptively faster in one PS task. Accuracy showed a similarly mixed pattern: only one task clearly favoured the chatbot, whereas the remaining contrasts were inconclusive ---and given the small sample, these contrasts should not be read as evidence of equivalence. They nevertheless suggest that conversational advantage was more clearly visible in participants' accounts of interactional effort than in stable improvements in objective task outcomes. Theme 1 is therefore best interpreted as explaining \emph{why} the CA was often experienced as helpful, while the quantitative results indicate that this experience did not reliably translate into faster or more accurate performance across tasks.

Theme 3 further suggests that user-side literacy may act as a boundary condition on the comparative value of the two interfaces. Participants often framed successful use of both interfaces as conditional on already knowing how to operate them, suggesting that the issue was not only interface learnability. It may also depend on pre-existing individual skills, including higher-order skills in the case of conversational interaction, such as the ability to formulate a clear and effective question. This is relevant because LLM-based CUIs may reduce demands on some forms of literacy, such as visualization and data literacy, which are central to efficient dashboard use \citep{wakeling_graph_2015}, while introducing other demands, such as prompt formulation and the ability to articulate one's reasoning clearly, as recently suggested by \citet{koloski_data_2025}. This interpretation is also consistent with \citet{liu_conversational_2024}, who suggest that a CUI can be advantageous for sporadic use, while a GUI may better support skill growth and therefore long-term use.

\subsection{Task complexity reshape interface advantages}
When analyzing our findings in relation to \textbf{RQ3}, the clearest pattern was that PS tasks were more demanding than IR tasks across both interfaces, as expected. They consistently took longer and also tended to elicit higher MWL. This supports the validity of our complexity manipulation based on Wood's formulation \citep{wood_task_1986}, operationalized here mainly through the greater number of information cues that had to be considered and coordinated. In this sense, the shift from IR to PS produced the expected increase in task demands.

More importantly, this increase in demand did not produce a stable shift in objective performance in favour of either the CA or the dashboard. Although the chatbot showed advantages in some tasks, these advantages did not generalize uniformly as demands increased; nor did the dashboard emerge as consistently superior in the more demanding conditions. In other words, the manipulation increased task burden, especially in terms of completion time and partly in subjective MWL, but it did not reveal a clear crossover or stable interface dominance. This is important because it suggests that greater task demands did not simply magnify the strengths of one interface while exposing the weaknesses of the other. This pattern is also compatible with \citet{bacic_task-representation_2018}, who suggests that the relation between perceived cognitive effort and objective performance does not remain constant across task-demand levels. 

Qualitative findings help refine this interpretation. Rather than describing one interface as categorically preferable for more demanding tasks, participants often articulated a more conditional and strategic logic of use, captured in Theme 5. Under greater complexity or uncertainty, some participants emphasized the value of overview and exploration, whereas others valued the CA's ability to assemble or summarize information that would otherwise require repetitive manual work. This suggests that, as task demands increased, the relevant question was often not which interface was globally superior, but which kind of support was most useful at a given stage of the task. From a design perspective, this indicates that LLM-based CUIs should not be treated as one-size-fits-all replacements for traditional GUIs. The results instead point to a broader interaction-design problem: identifying which style, or combination of styles, best fits the task and context of use.

\subsection{Useful, yes, but I wouldn't rely on it}
In relation to \textbf{RQ4}, one of the clearest findings is that perceived usefulness and confidence did not move together. The CUI was generally considered more useful than the GUI, but this was not accompanied by greater confidence in the CA. Confidence ratings did not favour the chatbot and, when confidence was made the explicit criterion for intended use, preference shifted toward the dashboard. This divergence suggests that the perceived convenience of conversational interaction did not automatically translate into an intention to rely on it.

Qualitative findings, particularly Theme 2, help explain this pattern. Participants often treated reliance on chatbot outputs as conditional rather than automatic, precisely because the answer was easy to obtain but harder to inspect. In this sense, the issue was not simply whether the chatbot was right or wrong, but whether users could understand the basis on which the answer had been produced and whether they could defend it as a legitimate basis for decision making. This helps explain why some participants re-asked questions to probe consistency, while others explicitly required technical conditions, such as the presence of RAG, as prerequisites for reliance. By contrast, the dashboard was often treated as more directly grounded in inspectable data, even when this required more effort to navigate and interpret.

A related finding is that participants often used chatbot outputs to complete the task while still expressing doubts about their accuracy or grounding. This suggests a form of provisional reliance rather than unconditional confidence. The source of low confidence also appeared to differ across interfaces. When participants expressed low confidence in the dashboard, this was often framed as a problem of their own unfamiliarity or lack of skill (Theme 3), and sometimes as a consequence of usability issues (Theme 4). Instead, when they expressed low confidence in the chatbot, it was more often framed as a problem of the system itself, especially its opacity and possible inconsistency (Theme 2), and only sometimes as a problem of the user's ability to formulate or articulate the request (Theme 3). This asymmetry is compatible with work on appropriate reliance in automation, which suggests that reliance depends on whether users feel able to calibrate when and how system's assistance should be followed \citep{lee_trust_2004}.

Taken together, these findings suggest that the CUI was often preferred as a useful and efficient interface, but not as an unquestioned basis for confident action. Participants expressed greater willingness to use the LLM-based CUI than the GUI in several circumstances, yet this preference was frequently qualified by the context of use. This pattern contrasts with work showing that direct-answer chatbot formats can increase behavioural reliance, particularly under time pressure \citep{spatola_efficiency-accountability_2024}. In our study, however, the convenience of direct answers did not automatically translate into reliance intentions, that is, participants’ stated willingness to rely on the chatbot’s outputs \citep{lee_trust_2004, jessup_closer_2024}, since they often qualified that willingness in terms of inspectability, verification, and direct access to underlying data, even while valuing its efficiency. However, this pattern should be interpreted cautiously, given that the sample consisted mainly of Computer Science students and therefore included participants who were likely to have relatively high familiarity with chatbot interaction and its limitations.

\subsection{Limitations}
This study is not without limitations. First, although the qualitative analysis followed a structured and documented procedure, the use of LLMs as coding assistants in thematic analysis remains methodologically unsettled. Recent work suggests that LLMs can support coding and theme development, but also highlights issues related to prompt sensitivity, output instability, hallucination, context-window constraints, and the need for human validation \citep{nguyen-trung_chatgpt_2025, wen_leveraging_2026}. In our case, this procedure was adopted partly to reduce dependence on the idiosyncratic interpretation of a single human coder, but it may also have introduced model-specific biases. To partially mitigate this risk, multiple models were used; however, this does not remove the need to interpret the qualitative findings as the outcome of a human-LLM analytic workflow.

Second, the sample was small and composed of students used as proxies for early-stage industrial decision makers. This choice was appropriate for an exploratory laboratory study, but limits generalization to experienced industrial decision makers, who may differ in domain knowledge, dashboard or data literacy, accountability pressures, and expectations toward AI-based tools.

Third, the experiment compared one specific dashboard with one specific chatbot implementation in a constrained manufacturing scenario. The observed differences may therefore reflect not only interaction style, but also implementation-specific design choices, usability limitations, and data-access capabilities of the particular systems tested.

Fourth, although IR tasks were deliberately administered before PS tasks, this fixed progression means that task complexity is partly confounded with sequence, increasing familiarity, and possible fatigue. This ordering was nevertheless a necessary design trade-off. Because PS tasks included IR subtasks, full counterbalancing would have introduced asymmetric carryover effects: participants completing PS tasks first would have acquired familiarity with retrieval operations before later performing the simpler IR tasks. Moreover, starting with the highest-complexity tasks would have been less representative of how users typically approach an unfamiliar interface. The IR-before-PS progression was therefore chosen to preserve a plausible learning trajectory, while acknowledging that sequence effects cannot be fully excluded.

Finally, the tasks were short, individual, and low-stakes compared with real industrial decision-making, where decisions are often collaborative, repeated over time, and embedded in organizational accountability structures. The ecological validity of the findings is therefore constrained, and the results should be read as evidence about situated interaction mechanisms under the tested conditions rather than as a definitive ranking of conversational interfaces and dashboards.

\section{Conclusions and Future Work}

This exploratory mixed-methods study investigated how an LLM-based CA and a graphical dashboard supported performance-management decision tasks in a simulated industrial context. The findings suggest that conversational interaction can reduce some of the burden associated with accessing and assembling information, particularly when users can express their needs as clear retrieval, filtering, comparison, or calculation requests. In these situations, the chatbot often reduced navigation and manual processing by returning a direct answer. However, this perceived reduction in effort did not amount to evidence of a consistent performance advantage across tasks.

Overall, rather than supporting the replacement of dashboards by conversational interfaces, the findings reveal complementary strengths and limitations. As task demands became more integrative, participants continued to value the dashboard as a persistent and inspectable workspace for obtaining an overview, comparing information, and verifying the basis of a decision. The chatbot could support calculation and synthesis, but its usefulness was conditioned by users' ability to formulate effective requests and by their confidence in the correctness and traceability of its outputs. Thus, the relative value of the two interaction styles appears to depend on the demands of the task and on the conditions under which users can interpret and rely on the resulting information.

Participants' accounts also identified potentially relevant user-side conditions, including dashboard familiarity, prompt formulation ability, domain knowledge, and comfort with textual or graphical representations. Because these factors emerged from participants' accounts rather than being tested as formal moderators, this study cannot establish whether or how systematically they shape workload, performance, or interface preference.

These findings should therefore be treated as exploratory evidence and as a basis for further testing. Given the small proxy sample and the constrained experimental setting, the study cannot establish whether the observed patterns generalize to industrial decision makers or hold systematically across different task demands and user capabilities. Nevertheless, it provides a rationale for a larger investigation designed to test these conditions more directly. Moreover, on the basis of the present evidence, the most plausible design direction is not the substitution of dashboards by chatbots, but the development of hybrid systems in which conversational agents reduce avoidable interaction costs while graphical interfaces preserve overview, transparency, and user control.


\section{Conflicts of interest}
One author holds a senior leadership role at Zerynth. All other authors declare no competing interests.

\section{Funding}
This work was supported by the National PhD Programme in Artificial Intelligence funded under Italy’s PNRR, NextGenerationEU (CUP I51J22000590007). Zerynth provided in-kind support by granting access to its dashboard platform and cloud-computing resources enabling access to GPT-4o via Microsoft Azure. The funders and sponsors had no role in the study design, data collection, data analysis, interpretation of results, manuscript preparation, or the decision to submit the article for publication.


\section*{Author contributions statement}
R.F. and A.S. conceived the study. R.F., S.C., A.S. and T.T. designed the experiment. R.F. developed the conversational system, conducted the experiment, collected and curated the data, and performed the qualitative analysis. R.F. and S.C. conducted the quantitative analyses. R.F., S.C., A.S. and T.T. interpreted the results. R.F. wrote the original draft of the manuscript. A.S. and T.T. critically revised the manuscript. D.M. provided access to the industrial resources and supervised the work. 

\section{Acknowledgments}
We thank Zerynth for providing access to the dashboard environment and computational resources used in this study.


\bibliographystyle{oup-abbrvnat}
\bibliography{reference}




\begin{appendices}
\section{Experimental Tasks}
\label{app:tasks}
This appendix reports the tasks participants were asked to perform and respond to in the first exploratory study.

\subsection*{Information Retrieval Tasks}

\paragraph{Task 1}
On which day in February did the factory consume the most energy?

\answerline

\paragraph{Task 2}
How much idle time did Laser Welding Machine 2 accumulate in the first week of March?

\answerline

\subsection*{Problem-Solving Tasks}

\paragraph{Task 3}
From February to March, your factory has noticed excessive energy waste from machines sitting idle for too long. Management is considering installing an automatic power-down feature on one cutting machine to reduce energy costs.

During a recent team meeting, one colleague pointed out that we should compare how long different machines sit idle, how much power they draw when idle, and how that translates into overall expenses.

Based on these observations, on which cutting machine would this feature have the greatest impact?

\answerline

\paragraph{Task 4}
An urgent order needs to be completed as quickly and efficiently as possible. The production process requires the following sequence of machines:

\begin{center}
\textit{Medium Capacity Cutting Machine}
\(\rightarrow\)
\textit{Assembly Machine}
\(\rightarrow\)
\textit{Testing Machine}
\end{center}

This should be evaluated in terms of average cycle time and performance during March. Select the machine sequence that should be assigned to this order.

\vspace{0.5\baselineskip}

\begin{center}
\begin{tabularx}{0.82\linewidth}{@{}Xccc@{}}
\toprule
\textbf{Machine type} & \textbf{1} & \textbf{2} & \textbf{3} \\
\midrule
Medium Capacity Cutting Machine & $\square$ & $\square$ & $\square$ \\
Assembly Machine                & $\square$ & $\square$ & $\square$ \\
Testing Machine                 & $\square$ & $\square$ & $\square$ \\
\bottomrule
\end{tabularx}
\end{center}

\section{Semi-Structured Interview and Post-Study Questionnaire}
\label{app:interview}

This appendix reports the semi-structured interview and post-study questionnaire administered after participants had completed all experimental tasks.

\subsection*{Overall Impressions}

```latex
\subsection*{Interface Usefulness and Confidence}

\noindent\textbf{Q1. How useful was the chatbot in completing the tasks?}

\vspace{0.3\baselineskip}
\begin{tabular}{@{}lll@{}}
\checkbox\ Very useful &
\checkbox\ Somewhat useful &
\checkbox\ Not useful
\end{tabular}

\vspace{0.5\baselineskip}
\textit{Follow-up:} Can you explain why you rated it that way?

\vspace{1.4\baselineskip}

\noindent\textbf{Q2. How useful was the dashboard in completing the tasks?}

\vspace{0.3\baselineskip}
\begin{tabular}{@{}lll@{}}
\checkbox\ Very useful &
\checkbox\ Somewhat useful &
\checkbox\ Not useful
\end{tabular}

\vspace{0.5\baselineskip}
\textit{Follow-up:} Can you explain why you rated it that way?

\vspace{1.4\baselineskip}

\noindent\textbf{Q3. Did you feel more confident using the chatbot or the dashboard?}

\vspace{0.3\baselineskip}
\begin{tabular}{@{}lll@{}}
\checkbox\ Chatbot &
\checkbox\ Dashboard &
\checkbox\ Both equally
\end{tabular}

\vspace{0.5\baselineskip}
\textit{Follow-up:} Why did you feel that way?

\subsection*{Preferences and Decision-Making}

\noindent\textbf{Q4. For which types of tasks would you prefer the chatbot or the dashboard?}

\vspace{0.5\baselineskip}

\begin{center}
\begin{tabularx}{0.82\linewidth}{@{}Xccc@{}}
\toprule
\textbf{Task type} & \textbf{Dashboard} & \textbf{Chatbot} & \textbf{Both equally} \\
\midrule
Information retrieval & \checkbox & \checkbox & \checkbox \\
Problem solving       & \checkbox & \checkbox & \checkbox \\
\bottomrule
\end{tabularx}
\end{center}

\vspace{1.4\baselineskip}

\noindent\textbf{Q5. If you had to use these tools again, under what circumstances would you choose the chatbot or the dashboard?}

\vspace{0.3\baselineskip}
For each circumstance, indicate the preferred option.

\vspace{0.5\baselineskip}

\begin{center}
\begin{tabularx}{\linewidth}{@{}Xccc@{}}
\toprule
\textbf{Circumstance} & \textbf{Dashboard} & \textbf{Chatbot} & \textbf{Both equally} \\
\midrule
When speed is important                        & \checkbox & \checkbox & \checkbox \\
When the task is complex                       & \checkbox & \checkbox & \checkbox \\
When I need to analyze a lot of data           & \checkbox & \checkbox & \checkbox \\
When I need direct answers to questions        & \checkbox & \checkbox & \checkbox \\
When I need high confidence in the information & \checkbox & \checkbox & \checkbox \\
\bottomrule
\end{tabularx}
\end{center}

\vspace{1.4\baselineskip}

\noindent\textbf{Q6. Do you have any other comments?}
```

\end{appendices}

\end{document}